\documentclass[prl,aps,reprint,superscriptaddress,amsmath]{revtex4-1}

\usepackage{bm}
\usepackage{graphicx}
\usepackage{dsfont}

\newcommand{\bra}[1]{\langle #1 | \,}
\newcommand{\ket}[1]{\, | #1 \rangle}
\newcommand{\braket}[2]{\langle #1 | #2 \rangle}
\newcommand{\expv}[1]{\langle #1 \rangle}
\newcommand{\bs}[1]{\boldsymbol{#1}}
\newcommand{\mr}[1]{\mathrm{#1}}
\newcommand{\mc}[1]{\mathcal{#1}}
\newcommand{\lra}{\leftrightarrow}
\newcommand{\om}{\omega}
\newcommand{\Om}{\Omega}
\newcommand{\ga}{\gamma}
\newcommand{\Ga}{\Gamma}
\newcommand{\de}{\delta}
\newcommand{\De}{\Delta}

\newcommand{\eps}{\epsilon}

\newcommand{\hlf}{\frac{1}{2}}
\newcommand{\sig}{\hat{\sigma}}

\begin{document}

\title{Deterministic free-space source of single photons using Rydberg atoms}

\author{David Petrosyan}
%\email{david.petrosyan@iesl.forth.gr}
\affiliation{Institute of Electronic Structure and Laser, FORTH,
GR-71110 Heraklion, Crete, Greece}

\author{Klaus M\o lmer}
\affiliation{Department of Physics and Astronomy, Aarhus University,
DK-8000 Aarhus C, Denmark}

\date{\today}

\begin{abstract}
%Preparing a single, isolated quantum system, such as an atom, molecule 
%or a quantum dot, in an excited state with high fidelity is routinely done 
%using coherent laser pulses, but deterministically converting this excitation
%into a single photon in a desired optical mode is technically demanding 
%and requires strong coupling of the atom to a resonant cavity. Here 
We propose an efficient free-space scheme to create single
photons in a well-defined spatio-temporal mode. 
To that end, we first prepare a single source atom in an excited Rydberg state. 
The source atom interacts with a large ensemble of ground-state atoms
via a laser-mediated dipole-dipole exchange interaction. 
Using an adiabatic passage with a chirped laser pulse, we produce 
a spatially-extended spin-wave of a single Rydberg excitation 
in the ensemble, accompanied 
by the transition of the source atom to another Rydberg state.
%In an experiment, the success probability of this operation can be further
%boosted by detecting the source atom in the corresponding Rydberg state,
%which heralds the spin-wave preparation in the atomic ensemble.
The collective atomic excitation can then be converted to a
propagating optical photon via a coherent coupling field. 
In contrast to previous approaches, our single-photon source does 
not rely on the strong coupling of a single emitter to a resonant cavity,
nor does it require the heralding of collective excitation or complete 
Rydberg blockade of multiple excitations in the atomic ensemble.
\end{abstract}

\maketitle

%\section{Introduction}

Single photons can serve as flying qubits for many important
applications, including all-optical quantum computation
and long-distance quantum communication and cryptography
\cite{Kimble2008,OBrien2009}.
Various sources of single photons are being explored; 
most of them use single emitters coupled to resonant cavities or waveguides
\cite{McKeever2004,Keller2004,Hijlkema2007,Stute2012,Reiserer2015,Matthiesen2012,Lodahl2015,Lodahl2017}.
Free-space schemes typically rely on the Duan-Lukin-Cirac-Zoller
protocol \cite{DLCZ2001} for the low-efficiency heralded preparation 
of a collective spin excitation of an atomic ensemble 
followed by its stimulated Raman conversion into
a photon \cite{EITrev2005,Hammerer2010,Sangouard2011}.
Creating a deterministic source of single photons
without requiring coupling to resonant optical
structures remains an outstanding challenge. Here we show how, in a
free-space setting, the remarkable properties of Rydberg atoms can be used
to map a single atomic excitation on a single photon emitted into
a well-defined spatial and temporal mode.

Atomic Rydberg states with high principal quantum numbers
$n \gg 1$ have long lifetimes $\tau \propto n^3$ and strong
electric dipole moments $\wp \propto n^2$ \cite{RydAtoms}.
The resulting long-range, resonant (exchange) and nonresonant
(dispersive or van der Waals) dipole-dipole interactions between
the atoms can suppress more than one Rydberg excitation within a certain
blockade distance \cite{Jaksch2000,Lukin2001,rydQIrev,rydDBrev}.
An ensemble of atoms in the blockaded volume can be viewed as 
an effective two-level Rydberg superatom 
\cite{Lukin2001,rydQIrev,Dudin2012,Ebert2015,Zeiher2015,DPGMN2014}.
A single collective excitation of the superatom can be created by resonant 
laser(s) and then converted to a photon in a well-defined spatio-temporal mode 
\cite{Saffman2002,Saffman2005,Pedersen2009,Nielsen2010,Miroshnychenko2013,Kuzmich13}.
%In fact, the single Rydberg excitation spin-wave in an atomic ensemble
%is equivalent to a coherently and reversibly stored single photon using
%stimulated Raman techniques \cite{Fleischhauer2000,EITrev2005,Gorshkov2007}.

There are several complications associated with the efficient
creation of a single coherent Rydberg excitation in an atomic ensemble
and its  deterministic conversion into a photon.
Creating only a single excitation requires a completely blockaded 
atomic ensemble. Efficient conversion of the excitation into a photon 
in a well-defined spatio-temporal mode requires a large optical depth.
Hence, the blockaded volume should accommodate many atoms, which
presumes strong, long-range, isotropic interactions.
The van der Waals interactions between  Rydberg excited atoms can 
be nearly isotropic \cite{Walker2008}, but a blockade range 
much beyond $10~\:\mu$m is difficult to achieve.
The large optical depth of the blockaded volume requires 
a high atomic density which, however, leads to a strong decoherence 
of the Rydberg-state electrons \cite{TPfau201346} and may involve
molecular resonances of Rydberg excited atoms \cite{Derevianko2015}.
Dipole-dipole interactions have a longer range, which allows 
using larger atomic ensembles with lower densities.
Compared to the van der Waals interaction, however, the dipole-dipole
interactions are ``softer'', leading to incomplete suppression of
multiple Rydberg excitations within the blockade distance \cite{Petrosyan2013}.
%which makes high-fidelity preparation of a single Rydberg excitation problematic.

In contrast, preparing a single, isolated atom in an excited state
with high fidelity is relatively easy. We propose an efficient free-space 
technique to convert this excitation into a single photon in a well-defined 
mode, without resorting to strong coupling of the atom to a single 
cavity mode \cite{McKeever2004,Keller2004,Hijlkema2007,Reiserer2015}.
Instead, we use long-range dipole-dipole exchange interactions between 
the Rydberg states of atoms to map the Rydberg excitation of the 
single ``source'' atom onto a collective Rydberg excitation of an 
ensemble of ``medium'' atoms. The mapping efficiency is boosted by 
the collectively enhanced coupling of the source atom to many medium atoms, 
but a complete Rydberg blockade or strong interactions among the medium 
atoms are not required. Subsequently, using a coupling laser pulse, 
the collective excitation of the ensemble of medium atoms having a large 
optical depth can be converted into a single photon propagating in a 
phase-matched direction. 
%In an experiment, the successful conversion can be 
%verified by detecting the source atom in the corresponding Rydberg 
%state. %which heralds the spin-wave preparation in the atomic ensemble.

We note a closely related recent proposal to realize a chiral light-atom 
interface by transferring the state of a single atom to a spatially ordered 
array of atoms which acts as a phased-array optical antenna for photon 
emission into a well-defined spatial mode \cite{Grankin2018}.

%We note that dipole-dipole exchange interactions between Rydberg
%states of atoms were recently used to demonstrate symmetry-protected
%collisions between propagating and stored photons in a Rydberg EIT
%medium \cite{Thompson2017}, and were explored for realizing
%dipolar exchange induced transparency in an atomic ensemble
%\cite{Petrosyan2017}.

%%%%%%%%%%FIGURE%%%%%%%%%%%%
\begin{figure}[t]
\centerline{\includegraphics[width=7cm]{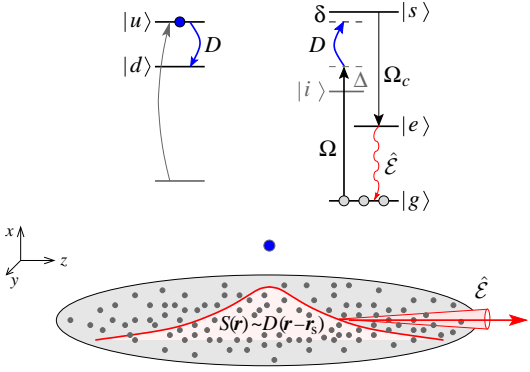}}
\caption{%(color online)
Schematics of the system.
A single ``source'' atom is initially prepared in the Rydberg state $\ket{u}$
(top left).
The transition $\ket{u} \to \ket{d}$ of the source atom is coupled
non-resonantly to the Rydberg transition $\ket{i} \to \ket{s}$ of
the ``medium'' atoms (top right) with the dipole-dipole exchange 
interaction $D$.
All medium atoms are initially in the ground state $\ket{g}$.
A laser pulse couples $\ket{g}$ to the intermediate Rydberg
state $\ket{i}$ with Rabi frequency $\Omega$ and detuning
$\Delta \gg |\Omega|,D$. Together with the dipole-dipole exchange
$\ket{i} \ket{u} \to \ket{s} \ket{d}$, this leads to the transition 
$\ket{g} \to \ket{s}$ of the medium atoms detuned by $\delta$.
The resulting single Rydberg excitation of the medium atoms has the
spatial amplitude profile $S(\bs{r}) \propto D(\bs{r} - \bs{r}_s)$.
The stored excitation can be converted to a propagating photon $\mc{E}$ 
by applying the control field $\Omega_c$ on the transition from $\ket{s}$ 
to the electronically excited state $\ket{e}$.}
\label{fig:scheme}
\end{figure}
%%%%%%%%%%%%%%%%%%%%%%%%%%%%%%%%%%%%%%%

%\section{The atomic system}

Consider the system shown schematically in Fig.~\ref{fig:scheme}.
We assume that a single source atom, having a strong dipole-allowed
microwave transition with frequency $\omega_{ud}$ between the Rydberg states
$\ket{u}$ and $\ket{d}$, is trapped in a well-defined spatial location 
that can be addressed with focused lasers. The source atom can 
be transferred from the ground state to the excited state $\ket{u}$ with unit
probability (see Fig.~\ref{fig:scheme} top left), by either a resonant laser
$\pi$-pulse or via an adiabatic transfer with a single chirped or a pair of
delayed laser pulses \cite{Bergmann1998,Vitanov2001}.

Consider next an ensemble of $N \gg 1$ medium atoms. The relevant states
of the atoms are the ground state $\ket{g}$, a lower electronically excited
state $\ket{e}$ and a pair of highly excited Rydberg states $\ket{i}$ and
$\ket{s}$ having a strong dipole-allowed transition with frequency $\omega_{si}$
(see Fig.~\ref{fig:scheme} top right).
A spatially uniform laser field couples non-resonantly 
the ground state $\ket{g}$ to the Rydberg state $\ket{i}$ with 
time-dependent Rabi frequency $\Omega$ and large detuning 
$\Delta \equiv \omega - \omega_{ig} \gg |\Omega|$.
The medium atoms interact with the source atom via the dipole-dipole 
exchange $\ket{i}  \ket{u} \lra \ket{s} \ket{d}$.
To be specific, we assume that the corresponding dipole matrix elements
of the medium and source atoms, $\bs{\wp}_{si}$ and $\bs{\wp}_{du}$, are 
along the $y$ direction.
% and we neglect angular corrections arising 
%from degeneracy of Rydberg Zeeman substates \cite{Walker2008}.
The exchange interaction strength is then
$D(\bs{R}) = \frac{C_3}{|\bs{R}|^3} (1 - 3 \cos^2 \vartheta)$, where
$C_3 = \frac{\wp_{si}  \wp_{du}}{4 \pi \eps_0 \hbar}$, 
$\bs{R} \equiv  (\bs{r}  -\bs{r}_\mr{s})$ is the relative position vector 
between an atom at position $\bs{r}$ and a source atom at $\bs{r}_\mr{s}$,
and $\vartheta$ is the angle between $\bs{R}$ and $\bs{\wp}_{si,du}$ 
($y$ direction). %, i.e., $\cos \vartheta = (y-y_\mr{s})/R$.
We assume a large frequency mismatch 
$\Delta_\mr{sa} \equiv \omega_{ud} - \omega_{si} \simeq - \Delta$,
$|\Delta_\mr{sa}| \gg D(\bs{R}) \; \forall \; \bs{R}$, which requires
positioning the source atom outside the volume containing the 
medium atoms (see below).

In the frame rotating with the frequencies $\omega$ and $\omega_{ud}$, the
Hamiltonian for the system is
\begin{eqnarray}
H_1/\hbar &=& - \sum_{j=1}^N
\big\{ \Delta \ket{i}_j\bra{i} + \delta \ket{s}_j\bra{s} \otimes 
\ket{d}_\mr{s} \bra{d}
\nonumber \\ & & \qquad
+ (\Omega e^{i \bs{k}_0 \cdot \bs{r}_j} \ket{i}_j\bra{g} + \mr{H.c.})
\nonumber \\ & & \qquad
- [D(\bs{R}_j) \ket{s}_j\bra{i} \otimes \ket{d}_\mr{s} \bra{u} 
+ \mathrm{H.c.}] \big\} , \label{eq:Ham1full}
\end{eqnarray}
where the index $j$ enumerates the medium atoms at positions $\bs{r}_j$,
$\bs{k}_0 \parallel \bs{e}_z$ is the wave vector of the laser field, and
$\delta \equiv \Delta + \Delta_\mr{sa} = \omega + \omega_{ud} - \omega_{sg}$
is the detuning of the product state  $\ket{s} \ket{d}$. 
Since the intermediate Rydberg level $\ket{i}$ is strongly detuned,
$\Delta \simeq - \Delta_\mr{sa} \gg |\Om|,D,|\delta|$, we can eliminate
it adiabatically. We then obtain an effective Hamiltonian
\begin{eqnarray}
\tilde{H}_1/\hbar &=& - \sum_j 
\big[\tilde{\delta}_j \ket{s}_j\bra{s} \otimes \ket{d}_\mr{s} \bra{d}
\nonumber \\ & & \qquad
+ (\tilde{D}_j e^{i \bs{k}_0 \cdot \bs{r}_j} \ket{s}_j\bra{g} \otimes \ket{d}_s \bra{u}
+ \mr{H.c.}) \big], \;\;
\label{eq:Ham1eff}
\end{eqnarray}
where $\tilde{\delta}_j \equiv \tilde{\delta}(\bs{R}_j)
= \delta + \frac{|\Omega|^2 - |D(\bs{R}_j)|^2}{\Delta}$ 
is the shifted detuning of $\ket{s}_j \ket{d}$ 
and $\tilde{D}_j \equiv \tilde{D}(\bs{R}_j)
= - \frac{D(\bs{R}_j) \Omega} {\Delta}$ is the second-order coupling
between $\ket{g}_j\ket{u}$ and $\ket{s}_j \ket{d}$.
$\tilde{\delta}(\bs{R})$ has a weak position dependence stemming from
the level shift $\frac{|D(\bs{R})|^2}{\Delta}$ of $\ket{s}$ due to 
the non-resonant dipole-dipole coupling, while the level shift 
$\frac{|\Omega|^2}{\Delta}$ of $\ket{g}$ is uniform.

%\subsection{Preparation of collective Rydberg excitation}

Let us for the moment neglect the (weak) spatial dependence of $\tilde{\delta}$,
i.e., assume that all the medium atoms have the same $\ket{g} \to \ket{s}$ 
transition frequency. Since by flipping the source atom
$\ket{u} \to \ket{d}$ we can create at most one Rydberg excitation
in the medium, we introduce the ensemble ground state
$\ket{G} \equiv \ket{g_1,g_2,\ldots,g_N}$
and a single collective Rydberg excitation state
$\ket{S} \equiv \frac{1}{\bar{D}} \sum_{j=1}^N
\tilde{D}_j e^{i \bs{k}_0 \cdot \bs{r}_j} \ket{s_j}$
($\ket{s_j} \equiv \ket{g_1,g_2, \ldots, s_j, \ldots, g_N}$) 
with $\bar{D} \equiv \big( \sum_j^N |\tilde{D}_j|^2 \big)^{1/2}$.
The Hamiltonian (\ref{eq:Ham1eff}) then reduces to that for a two-level system,
\begin{equation}
\tilde{H}_1 = - \hbar \left(\begin{array}{cc}
              0 & \bar{D} \\
              \bar{D} & \tilde{\delta}
        \end{array}\right)     \label{eq:Ham2ls}
\end{equation}
in the basis of states $\{\ket{G,u}, \ket{S,d} \}$.
The eigenstates and corresponding eigenvalues of this Hamiltonian are
$\ket{\pm} = \big [\mp \lambda_{\mp} \ket{G,u} \pm \bar{D} \ket{S,d} \big]
/\sqrt{\lambda_{\mp}^2 + \bar{D}^2}$ and 
$\lambda_{\pm} = \big[\tilde{\delta} \pm \sqrt{\tilde{\delta}^2 +4 \bar{D}^2 }
\big]/2$.
In the limit of large negative detuning $- \tilde{\delta} \gg \bar{D}$,
$\ket{+} \simeq \ket{G,u}$ with $\lambda_{+} \simeq 0$ 
coincides with the ensemble ground state. 
In the opposite limit of large positive detuning
$\tilde{\delta} \gg \bar{D}$, $\ket{+} \simeq \ket{S,d}$
with $\lambda_{+} \simeq \tilde{\delta}$ corresponds to the 
collective Rydberg excited state of the ensemble.
We can thus use the adiabatic passage \cite{Vitanov2001}
to convert the system from the initial ground state $\ket{G,u}$ to the Rydberg
excited state $\ket{S,d}$, by applying a laser pulse with a chirped frequency
$\omega = \omega_0 + \alpha (t-t_0)$, such that
$\tilde{\delta} = \alpha (t-t_0)$ is large and negative at early times
$t < t_0$ and is large and positive at later times $t > t_0$.
Provided the chirp rate satisfies $\alpha < \bar{D}^2$, the probability 
$P_{\ket{+} \to \ket{-}} =e^{-2 \pi \bar{D}^2/\alpha}$ of nonadiabatic Landau-Zener 
transition to the eigenstate $\ket{-}$ will be small \cite{Vitanov2001}. 
Since the coupling $\bar{D}$ is collectively enhanced by the 
number $N \gg 1$ of the medium atoms interacting with the source atom, 
we can use high chirp rates $\alpha$ to prepare the system with nearly 
unit probability in state $\ket{S,d}$. Moreover, we can verify the 
successful preparation of the medium in the collective state $\ket{S}$ 
by detecting the source atom in state $\ket{d}$ \cite{SM}.
We note that the adiabatic preparation of spatially ordered Rydberg excitations 
of atoms in a lattice using chirped laser pulses was studied in Refs.
\cite{Pohl2010,Schachenmayer2010,Schauss2015,Petrosyan2016,Malinovskaya2017}.

%%%%%%%%%%FIGURE%%%%%%%%%%%%
\begin{figure}[t]
\centerline{\includegraphics[width=8.7cm]{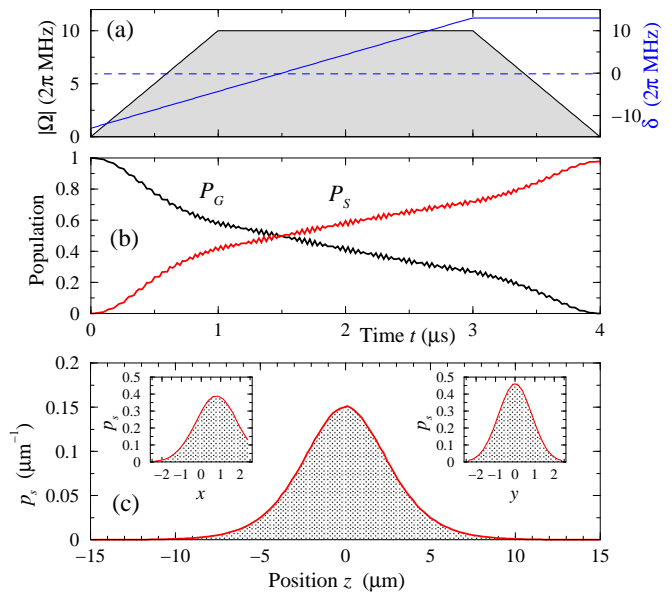}}
\caption{%(color online)
Preparation of a single collective Rydberg excitation of an atomic ensemble.
(a) Time dependence of Rabi frequency $\Omega$ (left vertical axis) 
and two-photon detuning $\delta$ (right vertical axis) of the driving laser.
(b) Dynamics of populations $P_G$ and $P_S$ of the
collective ground and single Rydberg excitation states of the atoms.
(c) Final spatial distribution of the Rydberg excitation 
in an elongated volume containing the atoms: 
The main panel shows the density of excitation $p_s(z)$ along
the longer $z$ axis of the volume (integrated over the transverse $x$ and $y$
directions), while the left and right insets show the densities
$p_s(x)$ and $p_s(y)$ along $x$ and $y$.
We average over $2000$ independent realizations of
the ensemble of $N=1000$ atoms with the peak density 
$\rho_{\max} \simeq 10 \:\mu\mbox{m}^{-3}$ and Gaussian  
distribution with $\sigma_{x,y} =1 \:\mu$m and $\sigma_{z} =6 \:\mu$m.
The source atom is placed
%initially in state $\ket{u}$ is placed 
%$7 \sigma_x$ away from the center of the cloud 
at $x_\mr{s} = 7\:\mu$m, $y_\mr{s}, z_\mr{s}=0$.
In the simulations, we take the peak Rabi frequency
$\Omega_{\max} = 2\pi \times 10 \:$MHz, the Rydberg state 
decay $\Gamma_s = 10\:$kHz and dephasing $\gamma_{sg} = 10\:$kHz,
and we use the atomic parameters corresponding to a Cs
source atom with $\ket{u} \equiv 70P_{3/2}$ and $\ket{d} \equiv 70S_{1/2}$, 
and Rb medium atoms with $\ket{i} \equiv 58P_{3/2}$ and 
$\ket{s} \equiv 57 D_{5/2}$ \cite{SM}. 
With the intermediate state detuning $\Delta = 9.4 \Omega_{\max}$
and the dipole-dipole interaction coefficient  
$C_3 = 11.7\:\mr{GHz}\,\mu\mathrm{m}^3$, the medium atoms 
near the cloud center experience the strongest interaction 
$D \simeq  2\pi \times 5.4\:$MHz. The corresponding second-order 
transition rate is $\tilde{D}_{\max} \simeq  2\pi \times 0.6 \:$MHz, 
while the collective coupling rate is $\bar{D} \simeq  2\pi \times 13 \:$MHz.}
\label{fig:OdPtxyzgr}
\end{figure}
%%%%%%%%%%%%%%%%%%%%%%%%%%%%%%%%%%%%%%%

We have performed exact numerical simulations of the dynamics 
of the system using realistic atomic parameters \cite{SM}. 
We place $N \gg 1$ ground-state $\ket{g}_j$ atoms in an elongated volume
at random positions $\bs{r}_j$ normally distributed around the origin,
$x,y,z=0$, with standard deviations $\sigma_{z} > \sigma_{x,y}$.
The source atom initially in state $\ket{u}$ is placed at a position 
$\bs{r}_\mr{s}$ close to the center of the longitudinal extent of the volume
and well outside its transverse width.
%$y_\mr{s},z_\mr{s} = 0$, and several microns outside of it, 
%$x_\mr{s} > 3 \sigma_x$, so that $D(\bs{R}_j)$ has the same sign
%for all the atoms $j$ in the volume [$\cos^2 (\vartheta) < 1/3$].
We apply to the atoms a pulsed and chirped laser field with
the Rabi frequency $\Omega$ and two-photon detuning $\delta$ 
shown in Fig.~\ref{fig:OdPtxyzgr}(a).
We simulate the evolution of the state vector of the system 
$\ket{\Psi_1} = c_0 \ket{G} \otimes \ket{u} + 
\sum_{j=1}^N  c_j e^{i \bs{k}_0 \cdot \bs{r}_j} \ket{s_j} \otimes \ket{d}$,
taking into account the decay and dephasing of the Rydberg states
\cite{SM}. The resulting dynamics of populations $P_G \equiv |c_0|^2$ and
$P_S \equiv \sum_j |c_j|^2$ are shown in Fig.~\ref{fig:OdPtxyzgr}(b).
We obtain a single collective Rydberg excitation $\ket{S}$ of the
atomic ensemble with high probability $P_S \gtrsim 0.977$, which is slightly
smaller than unity mainly due to the decay and dephasing. 
In Fig.~\ref{fig:OdPtxyzgr}(c) we show the final spatial 
distribution of the Rydberg excitation density 
$p_s (\bs{r}) \propto |\tilde{D}(\bs{r}-\bs{r}_\mr{s})|^2$
which follows the dipole-dipole interaction strength.
Our simulations verify that we can reliably prepare a spin wave of 
single collective Rydberg excitation $\ket{S}$ with the spatial wave function
\[
S(\bs{r}) 
% = \sum_j^N u_j(\bs{r}) c_j e^{i \bs{k}_0 \cdot \bs{r}_j} 
\simeq - \frac{\sqrt{\rho(\bs{r})} D(\bs{r} -\bs{r}_\mr{s})}
{\sqrt{\int dr^3 \rho(\bs{r}) D^2(\bs{r} -\bs{r}_\mr{s})}} 
e^{i \bs{k}_0 \cdot \bs{r}} .
\]
%where 
%$u_j(\bs{r}) = 1/\sqrt{V_j}$ within the volume 
%$V_j \simeq 1/\rho(\bs{r}_j)$ allotted to atom $j$ and $u_j(\bs{r}) = 0$ 
%otherwise, such that $\int dr^3 u_i(\bs{r})u_j(\bs{r}) = \delta_{ij}$, while 
%$\mathcal{N} = \int dr^3 \rho(\bs{r}) D^2(\bs{r} -\bs{r}_\mr{s})$.

%\subsection{Conversion of Rydberg excitation to a photon}

\begin{figure}[t]
\centerline{\includegraphics[width=8.7cm]{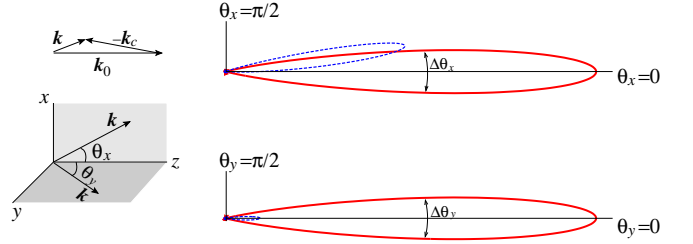}}
\caption{%(color online)
Angular probability distribution of the photon emitted by the atomic
medium. In the upper polar plot, the polar angle $\theta_x$ is varied
in the $x-z$ plane (azimuth $\varphi =0,\pi$);
in the lower plot, the polar angle $\theta_y$ is varied in the
$y-z$ plane (azimuth $\varphi =\pi/2,3\pi/2$).
The red solid line corresponds to the control field wave vector
$\bs{k}_c \parallel \bs{k}_0$ and the blue dashed line to a small 
inclination $\bs{k}_c \protect\angle \bs{k}_0 = 0.04 \pi$.
In the collinear geometry (red solid line), the resulting angular
width (FWHM) of the emitted radiation is
$\Delta \theta_x \simeq 0.07 \pi$ and $\Delta \theta_y \simeq 0.068 \pi$,
and the total probability of radiation emitted into the phase-matched
direction $z$ within the solid angle
$\Delta \Omega = 2 \pi (1- \cos \Delta \theta)$, with
$\Delta \theta \simeq 0.07 \pi$, is $\sim 0.74$.
The plots are obtained from a single realization of random positions 
of Rb atoms with the parameters in Fig.~\ref{fig:OdPtxyzgr}, but 
the quantum state of the emitted radiation $\ket{\psi_\mr{ph}}$ is highly
reproducible for different realizations ($m,m'$) of the atomic ensemble,
$|\braket{\psi_\mr{ph}^{(m)}}{\psi_\mr{ph}^{(m')}}| \gtrsim 0.96 P_S$. }
\label{fig:Pkang}
\end{figure}

Consider now the conversion of the collective Rydberg excitation of
the atomic medium into a photon. To that end, we use
a control laser field with wave vector $\bs{k}_c$ and frequency
$\omega_c = c k_c$ acting resonantly on the transition
$\ket{s} \to \ket{e}$ with the Rabi frequency $\Omega_c$.
The atomic transition $\ket{e} \to \ket{g}$ is coupled with 
strengths $g_{\bs{k},\sigma}$ to the quantized radiation field modes 
$\hat{a}_{\bs{k},\sigma}$ characterized by the wave vectors $\bs{k}$, 
polarization $\sigma$ and frequencies $\omega_k =c k$.
We take $\bs{k}_c \parallel \bs{k}_0$ so as to achieve resonant emission
of the photon in the phase-matched direction,
$\bs{k} = \bs{k}_0 - \bs{k}_c  \parallel \bs{e}_z$.
%see Fig.~\ref{fig:Pkang} and discussion below.
The frequency and wave number of the Rydberg microwave transition 
can be neglected in comparison with those of the optical transitions. 
In the frame rotating with frequencies $\omega_{rg}$ and 
$\omega_c= \omega_{re}$ (interaction picture), the Hamiltonian reads
\begin{eqnarray}
H_2/\hbar &=& - \sum_{j=1}^N \Big[
\sum_{\bs{k},\sigma}  g_{\bs{k},\sigma}  \hat{a}_{\bs{k},\sigma}
e^{i \bs{k} \cdot \bs{r}_j} e^{-i (\omega_k - \omega_{eg}) t} \ket{e}_j\bra{g}
\nonumber \\ & & \qquad \quad
+ \Omega_c e^{i \bs{k}_c \cdot \bs{r}_j} \ket{s}_j\bra{e} + \mathrm{H.c} \Big] .
\label{eq:Ham2}
\end{eqnarray}
The state vector of the system can be expanded as
$\ket{\Psi_2} = \sum_{j=1}^N \big[ c_j e^{i \bs{k}_0 \cdot \bs{r}_j} \ket{s_j}
+ b_j \ket{e_j} \big] \otimes \ket{0} 
+ \sum_{\bs{k},\sigma} a_{\bs{k},\sigma} \ket{G} \otimes \ket{1_{\bs{k},\sigma}}$,
where $\ket{e_j} \equiv \ket{g_1,g_2, \ldots, e_j, \ldots, g_N}$ and
$\ket{1_{\bs{k},\sigma}} \equiv \hat{a}^{\dagger}_{\bs{k},\sigma}  \ket{0}$
denotes the state of the radiation field with one photon in mode
$\bs{k},\sigma$.
Using the standard procedure \cite{SM}, 
%from the equations for the amplitudes $\hat{a}_{\bs{k},\sigma}$ and $b_j$ 
we obtain that the atomic state $\ket{e}$ spontaneously decays 
with rate $\Gamma_e$. Assuming $\Gamma_e \gg |\Omega_c|$ and 
eliminating $b_j$ leads to the solution for the amplitudes 
of photon emission into states $\ket{1_{\bs{k}}}$, 
\begin{equation}
a_{\bs{k}} (t) = - \tilde{g}_k(t) \sum_j c_j(0)
e^{i (\bs{k}_0 - \bs{k}_c -\bs{k}) \cdot \bs{r}_j} , \label{eq:aksol} 
\end{equation}
where 
$\tilde{g}_k(t) \equiv g_k \int_0^t dt'\frac{\Omega_c^*(t')}{\Gamma_e/2}
e^{i (\omega_k - \omega_{eg}) t'} e^{- \int_0^{t'} d t'' \frac{|\Omega_c(t'')|^2}{\Gamma_e/2} }$ 
and for simplicity we drop the polarization index $\sigma$ 
assuming scalar and isotropic emission by individual atoms.
In the case of a time-independent control field $\Omega_c$,
the dimensionless coupling reduces to $\tilde{g}_k \simeq 
\frac{g_k \Omega_c^*}{\Gamma_e(\omega_k - \omega_{eg})/2 + i |\Omega_c|^2}$
for $t \gg w^{-1}$.
The probability distribution of emitted photon $P_{\bs{k}} = |a_{\bs{k}}|^2$
is thus strongly peaked at frequency $\omega_k = \omega_{eg}$ with a narrow
linewidth $w = \frac{|\Omega_c|^2}{\Gamma_e/2}$, and
has a narrow angular distribution around
$\bs{k} = \bs{k}_0 - \bs{k}_c  \parallel \bs{e}_z$
as shown in Fig.~\ref{fig:Pkang}.
%Even though the positions $\{\bs{r}_j \}$ of the atoms within 
%the trapping volume vary between different realizations of the ensemble, 
Our simulations reveal that the state of the emitted photon 
$\ket{\psi_{\mr{ph}}} = \sum_{\bs{k}} a_{\bs{k}} \ket{1_{\bs{k}}}$
is largely insensitive to the microscopic details of various realizations
of the atomic ensemble.
Note that the efficient conversion of the atomic excitation into a photon 
requires a collinear geometry $\bs{k}_c \parallel \bs{k}_0$ for resonant 
emission at frequency $\omega_p = c |\bs{k}_0 - \bs{k}_c| \simeq \omega_{eg}$,
while even a small inclination $\bs{k}_c \protect\angle \bs{k}_0 \neq 0$
disturbs the phase matching in a spatially extended atomic medium
and reduces the probability of photon emission into the
well-defined spatial direction.
In the presence of the control field $\Omega_c$, the resonant photon 
propagating in the optically dense medium in the $z$ direction is subject 
to electromagnetically induced transparency (EIT) \cite{EITrev2005}, which 
suppresses photon reabsorption and scattering \cite{SM}.

%%%%%%%%%%FIGURE%%%%%%%%%%%%
%%%%%%%%%%%%%%%%%%%%%%%%%%%%%%%%%%%%%%%

%\subsection{Singe-step creation of the photon}

Similar to cavity QED schemes 
\cite{McKeever2004,Keller2004,Hijlkema2007,Reiserer2015} employing
stimulated Raman adiabatic passage \cite{Bergmann1998,Vitanov2001},
we can create a single photon directly, populating only virtually 
the Rydberg state $\ket{s}$.
To that end, we assume a constant control field $|\Omega_c| < \Gamma_e/2$
and small decay and two-photon detuning of the Rydberg state
$\Gamma_s,\tilde{\delta}_j \ll w$.
We then obtain a solution for the photon amplitudes as in Eq.~(\ref{eq:aksol}) 
with $c_j(0) \to -D_j/\Delta$ and
$\tilde{g}_k(t) \equiv g_k \int_0^t dt'\frac{\Omega(t)}{\Omega_c}
e^{i (\omega_k - \omega_{eg}) t'} e^{- \int_0^t dt'' \frac{\bar{D}^2(t'')}{w}}$ \cite{SM}.
Thus, with the source atom in state $\ket{u}$, the medium atoms in the
collective ground state $\ket{G}$, and the control field $\Omega_c \neq 0$,
by turning on the excitation laser $\Omega$ we produce a single photon
on the atomic transition $\ket{e} \to \ket{g}$.
This single-photon wave packet is emitted with high probability into the
direction of $\bs{k} \simeq \bs{k}_0 - \bs{k}_c  \parallel \bs{e}_z$,
while its temporal shape can be manipulated by the time dependence of
$\Omega(t)$. The emission of the optical photon is accompanied by the
transition of the source atom to state $\ket{d}$, which terminates the
conversion process, even if $\Omega \neq 0$.
To produce another photon, we have to reset the source atom to state $\ket{u}$.
%Hence, the source atom plays the role of a switch for the single photon
%production.

%\section{Discussion and Conclusions}

In summary, we have presented a new scheme for efficient 
single-photon production, controlled by a single source atom prepared 
in an appropriate Rydberg state and playing the role of a switch.
The dipole-dipole exchange interaction with the source atom enables 
single collective Rydberg excitation of the atomic ensemble without 
the requirement of a full blockade of the entire ensemble. 
Detailed numerical simulations with realistic experimental parameters
demonstrate that this excitation can be converted into a single photon 
emitted into a well-defined spatio-temporal mode with better 
than 70\% probability.
The probability that the photon is coherently emitted into the small
solid angle $\Delta \Omega \simeq 0.15\:$sr (see Fig. \ref{fig:Pkang})
is given approximately by $P_{\Delta \Omega} \simeq \eta N \Delta \Omega/4\pi$
\cite{Saffman2005}, where $\eta \simeq 0.6$ is the effective fraction
of the medium atoms participating in the collective Rydberg excitation
$\ket{S}$ (see Fig. \ref{fig:OdPtxyzgr}(c)).
%due to their dipole coupling with the source atom.  
This probability can be enhanced by increasing the atom number $N$, 
which may, however, lead to increased Rydberg state dephasing.

In our analysis, we assumed an ensemble of atoms at random positions
and with moderate density and neglected the vacuum field-mediated
interactions between the atoms on the optical transitions. 
Recently, Grankin {\it et al.} \cite{Grankin2018} have shown that imprinting 
an appropriate spatial amplitude and phase on an array of atoms with 
subwavelength spacing and coupled to a single atom in the same way as 
in our proposal can further enhance the photon emission probability 
into the predefined Gaussian (paraxial) mode of the radiation field,
which can be used for quantum state transfer between distant atoms
in free-space quantum networks.  
We finally note that our method to convert a single atomic excitation
to an optical photon can be used for microwave to optical conversion 
in hybrid quantum interfaces \cite{Kurizki2015,Gard2017}. 
For example, the source atom can be replaced by a superconducting qubit, 
which can strongly couple to the Rydberg states of the medium atoms by 
a microwave transition.
%Moreover, by exploiting sequences of coherent transitions among several
%internal states in the source atom or superconducting qubit, one may 
%produce trains of entangled pulses that are, e.g., resilient to loss 
%and dephasing during qubit transmission and quantum key distribution
%\cite{Nielsen2010}.

\begin{acknowledgments}
We thank Michael Fleischhauer and Thomas Pohl for useful discussions
and we are grateful to Mark Saffman for valuable advice and contributions.
We acknowledge support by the Villum Foundation (K.M.), 
the Alexander von Humboldt Foundation (D.P.) and 
the U.S. ARL-CDQI program through cooperative Agreement No. W911NF-15-2-0061.
\end{acknowledgments}

%%%%%%%%%%%%%%%%%%%%%%%%%%%%%%%%%%%%%
%%%%%%%%%%%%%%%%%%%%%%%%%%%%%%%%%%%%%

\newpage

\section{Supplementary Material}

In these notes, we present the details of derivation of the equations 
describing the spatially-extended single Rydberg excitation 
in the atomic ensemble and its subsequent conversion to 
a propagating optical photon. 
We also present possible choices of atoms and their Rydberg states 
and discuss the influence of multiple atomic excitations and transitions 
to other states on the performance of the single-photon source.

\subsection{Creation of a single collective Rydberg excitation in the atomic 
ensemble}
\label{ap:collRyd}

We consider the dynamics of the medium atoms coupled to the source atom 
and driven by a chirped laser field, as described in the main text.
The state vector of the combined system can be expanded as
\begin{equation}
\ket{\Psi_1} = c_0 \ket{G} \otimes \ket{u}
+ \sum_{j=1}^N  c_j e^{i \bs{k}_0 \cdot \bs{r}_j}
\ket{s_j} \otimes \ket{d} ,
\end{equation}
where $\ket{G} \equiv \ket{g_1,g_2,\ldots,g_N}$ and 
$\ket{s_j} \equiv \ket{g_1,g_2, \ldots, s_j, \ldots, g_N}$.
The time evolution is governed by the Schr\"odinger equation
$\partial_t \ket{\Psi_1} = -\frac{i}{\hbar} \tilde{H}_1\ket{\Psi_1}$
with the Hamiltonian (2) of the main text, which leads to the system
of coupled equations for the amplitude $c_0$ and the slowly varying 
in space amplitudes $c_j$:
\begin{subequations}
\label{eqs:c0cj}
\begin{eqnarray}
\partial_t c_0 &=& i \sum_{j=1}^N \tilde{D}_j^* c_j , \\
\partial_t c_j &=& i (\tilde{\delta}_j -\Gamma_s /2) c_j
+ i \tilde{D}_j \tilde{c}_0 .
\end{eqnarray}
\end{subequations}
We assume that the medium atoms in the Rydberg state $\ket{s}$ 
spontaneously decay with rate $\Gamma_s$ to other 
states $\ket{o} \neq \ket{g}$ which are decoupled from the laser 
and thereby the dynamics of the system. The atomic system is thus 
not closed, and the evolution is non-unitary, with a loss of norm 
of $\ket{\Psi_1}$ that ultimately reduces the photon emission probability.

We can also include the dephasing $\gamma_{sg}$ of the Rydberg state with 
respect to the ground state by randomly modulating in time the detunings
$\tilde{\delta}_j$ of individual atoms. 
Indeed, for any pair of atomic state $\ket{a}$ and $\ket{b}$, the dephasing
with rate $\gamma$ would result in the decay of the atomic coherence
$\rho_{ab} = c_a c_b^*$ as $\rho_{ab}(t) = \rho_{ab}(0) e^{-\gamma t}$.
In the differential equations for the amplitudes $c_a$ and $c_b$
of states $\ket{a}$ and $\ket{b}$, we can model this dephasing
by adding to the energy of state $\ket{b}$ (or to the detuning $\delta$)
a stochastic term $\varsigma(t)$ corresponding to a Gaussian random variable
with the variance $\expv{\varsigma^2} = \sigma^2$. It is then easy to verify
that in the coarse-grained integration of equations for $\dot{c}_a$
and $\dot{c}_b$ with small time steps $dt$, the variance should be
set to $\expv{\varsigma^2} = 2\gamma/dt$.

%%%%%%%%%%%%%%%%%%%%
\subsection{Conversion of the collective Rydberg excitation 
to an optical photon}
\label{ap:emitphoton}

We next consider the atomic ensemble prepared initially in state
$\ket{S} = \sum_{j=1}^N  c_j e^{i \bs{k}_0 \cdot \bs{r}_j} \ket{s_j}$.
As described by the Hamiltonian (4) of the main text, a
resonant control laser with the Rabi frequency $\Omega_c$ is applied 
to the the atomic transition $\ket{s} \to \ket{e}$, while the  
transition $\ket{e} \to \ket{g}$ is coupled with strengths 
$g_{\bs{k},\sigma} = \frac{\bs{\wp}_{eg} \cdot \bs{e}_{\bs{k},\sigma} }{\hbar}
\sqrt{\frac{\hbar \omega_k}{2 \eps_o V}}$ to the quantized radiation 
field modes within the quantization volume $V$.
We expand the state vector of the system as
\begin{eqnarray}
\ket{\Psi_2} &=& \sum_{j=1}^N \big[ c_j e^{i \bs{k}_0 \cdot \bs{r}_j} \ket{s_j}
+ b_j \ket{e_j} \big] \otimes \ket{0}
\nonumber \\ & &
+ \sum_{\bs{k},\sigma} a_{\bs{k},\sigma} \ket{G} \otimes \ket{1_{\bs{k},\sigma}} ,
 \end{eqnarray}
where $\ket{e_j} \equiv \ket{g_1,g_2, \ldots, e_j, \ldots, g_N}$ 
denotes the state with atom $j$ in the lower electronically excited state
$\ket{e}$ and $\ket{1_{\bs{k},\sigma}} \equiv \hat{a}^{\dagger}_{\bs{k},\sigma}  \ket{0}$
denotes the state of the radiation field with one photon in mode
$\bs{k},\sigma$. The resulting equations for the amplitudes are
\begin{subequations}
\begin{eqnarray}
\partial_t c_j &=& i \Omega_c  e^{i (\bs{k}_c -\bs{k}_0) \cdot \bs{r}_j} b_j , \\
\partial_t b_j &=& i \Omega_c^* e^{i (\bs{k}_0 -\bs{k}_c) \cdot \bs{r}_j} c_j
\nonumber \\ & &
+ i \sum_{\bs{k},\sigma}  g_{\bs{k},\sigma} e^{i \bs{k} \cdot \bs{r}_j} a_{\bs{k},\sigma}
e^{i (\omega_{eg} - \omega_k) t} ,  \\
\partial_t a_{\bs{k},\sigma} &=& i g_{\bs{k},\sigma}^* \sum_j
 e^{-i \bs{k} \cdot \bs{r}_j} b_j e^{i (\omega_k - \omega_{eg}) t} ,
\end{eqnarray}
\end{subequations}
with the initial conditions
$a_{\bs{k},\sigma}(0) = 0 \, \forall \, \bs{k},\sigma$,
$b_j (0) =0 \, \forall \, j$ and
$c_j (0)$ given by the solution of Eqs.~(\ref{eqs:c0cj}).
We rewrite the last equation for the photon amplitudes in the integral form,
\begin{equation}
a_{\bs{k},\sigma} (t) = i g_{\bs{k},\sigma}^* \sum_{j} e^{-i \bs{k} \cdot \bs{r}_j}
\int_0^t dt' b_j(t') e^{i (\omega_k - \omega_{eg}) t'} , \label{eq:akint}
\end{equation}
and substitute into the previous equation for the atomic amplitudes $b_j$.
We then obtain the usual spontaneous decay of the atomic state $\ket{e}$
with rate $\Gamma_e$, and the Lamb shift that can be incorporated into
$\omega_{eg}$ \cite{ScullyZubary1997,PLDP2007}.
We assume that the mean interatomic distance $\bar{r}_{ij}$ is large enough,
$k \bar{r}_{ij} = 2\pi \bar{r}_{ij}/\lambda > 1$, and neglect the
field--mediated interactions between the atoms
\cite{Miroshnychenko2013,Lehmberg1970,Thirunamachandran}.
The equations for the atomic amplitudes reduce to
\begin{subequations}
\label{eqs:cjbj}
\begin{eqnarray}
\partial_t c_j &=& i \Omega_c  e^{i (\bs{k}_c -\bs{k}_0) \cdot \bs{r}_j} b_j , \\
\partial_t b_j &=& - \hlf \Gamma_e b_j +
i \Omega_c^* e^{i (\bs{k}_0 -\bs{k}_c) \cdot \bs{r}_j} c_j  .
\end{eqnarray}
\end{subequations}
Assuming $\Gamma_e \gg |\Omega_c|$ and setting $\partial_t b_j =0 $, we
obtain
\[
b_j(t) = i c_j(0) e^{i (\bs{k}_0 -\bs{k}_c) \cdot \bs{r}_j}
\frac{\Omega_c^*(t)}{\Gamma_e/2}
e^{- \int_0^t dt' \frac{|\Omega_c(t')|^2}{\Gamma_e/2} }.
\]
The amplitudes for the emission of the photon into states $\ket{1_{\bs{k}}}$
are then
\begin{eqnarray}
a_{\bs{k}} (t) &=& - \tilde{g}_k(t) \sum_j c_j(0)
e^{i (\bs{k}_0 - \bs{k}_c -\bs{k}) \cdot \bs{r}_j} , \label{ap:eq:aksol} \\
\tilde{g}_k(t) & \equiv & g_k \int_0^t dt'\frac{\Omega_c^*(t')}{\Gamma_e/2}
e^{i (\omega_k - \omega_{eg}) t'} e^{- \int_0^{t'} d t'' \frac{|\Omega_c(t'')|^2}{\Gamma_e/2} },
\nonumber
\end{eqnarray}
where for simplicity we assume an isotropic atomic dipole, with
$g_{\bs{k},\sigma}  = g_k = \wp_{eg} \sqrt{\frac{\omega_k}{2 \hbar \eps_o V}}$
being a smooth function of frequency $\omega_k$.
In the case of a time-independent control field $\Omega_c$,
the dimensionless coupling $\tilde{g}_k$ reduces to
\begin{eqnarray*}
\tilde{g}_k(t) &=& \frac{g_k \Omega_c^*}{\Gamma_e/2}
\frac{ 1 - e^{i (\omega_k - \omega_{eg}) t} e^{- \frac{|\Omega_c|^2}{\Gamma_e/2} t} }
{(\omega_k - \omega_{eg}) + i \frac{|\Omega_c|^2}{\Gamma_e/2} } \\
&\to & \frac{g_k \Omega_c^*}{\Gamma_e(\omega_k - \omega_{eg})/2 + i |\Omega_c|^2}
\;\; \mbox{for} \;\; t \gg w^{-1},
\end{eqnarray*}
which is strongly peaked at frequency $\omega_k = \omega_{eg}$ with a narrow
linewidth $w = \frac{|\Omega_c|^2}{\Gamma_e/2}$.

\subsubsection{Electromagnetically Induced Transparency for the 
spin-wave to photon conversion}
\label{ap:EIT}

The single Rydberg excitation spin-wave of the atomic ensemble, 
which we convert with the resonant control field $\Omega_c$ to 
a single photon propagating in the phase-matched direction 
$\bs{k} = \bs{k}_0 - \bs{k}_c  \parallel \bs{e}_z$,
is equivalent to a coherently and reversibly stored single 
photon using electromagnetically induced transparency (EIT)
\cite{Fleischhauer2000,EITrev2005,Gorshkov2007}. Non-linear 
quantum optics using Rydberg EIT with ladder configuration of 
the atomic levels has recently been the subject of extensive research 
\cite{Murray2016,Firstenberg2016}. Here we are concerned only with 
the linear regime as we are dealing with a single atomic/photonic excitation.

We consider a one dimensional propagation and interaction of a quantum
field $\hat{\mathcal{E}}$ with the atomic medium on the transition
$\ket{g} \lra \ket{e}$, in the presence of the control field $\Omega_c$
resonantly driving the adjacent transition $\ket{e} \lra \ket{s}$ to 
the Rydberg state, $\omega_c=\omega_{se}$. 
The quantum field of carrier frequency $\omega_p \simeq \omega_{eg}$
and wave number $\omega_p/c$ is described by the slowly varying in time 
and space operator $\hat{\mathcal{E}}(z,t) 
= \frac{1}{\sqrt{L}} \sum_k \hat{a}_k e^{i(k-\omega_p/c)z} e^{i \om_p t}$,
where $L$ is the length of the quantization volume $V=AL$. The field 
operators obey the commutation relations 
$[\hat{\mc{E}}(z),\hat{\mc{E}}(z')] 
= [\hat{\mc{E}}^{\dag}(z),\hat{\mc{E}}^{\dag}(z')] = 0$ and
$[\hat{\mc{E}}(z),\hat{\mc{E}}^{\dag}(z')] = \de(z-z')$ which 
follow from the bosonic nature of operators $\hat{a}_k,\hat{a}^{\dag}_k$
for the individual longitudinal modes $k$. 
We assume that atom-field coupling strength 
$g = \wp_{eg} \sqrt{\frac{\omega_p}{2 \hbar \eps_o A}}$ is nearly
uniform within the relevant frequency bandwidth $w$. In the frame
rotating with frequencies $\om_p$ and $\omega_c$, the Hamiltonian
reads (cf. $H_2$ of Eq. (4) in the main text),
\begin{eqnarray}
H_2'/\hbar &=& -i c 
\int \! d z \, \hat{\mc{E}}^{\dag}(z) \, \partial_z \hat{\mc{E}}(z)
\nonumber \\ & & 
-  \int \! d z  \rho(z) \big[ \De_p \big( \sig_{ee}(z) + \sig_{ss} (z) \big) 
\nonumber \\ & & \qquad \quad
+ \big(g \hat{\mc{E}}(z) \sig_{eg} (z) + \Om_c  \sig_{se} (z) + 
\mathrm{H.c.}\big) \big]  , \quad \; \label{app:eq:Ham2}
\end{eqnarray}
where $\Delta_p = \om_p-\omega_{eg}$ is the detuning, 
$\rho(z)$ is the linear density of atoms, and we use the continuous 
atomic operators 
$\sig_{\mu\nu}(z) \equiv \frac{1}{N_z} \sum_i^{N_z} \ket{\mu}_i\bra{\nu}$
averaged over $N_z = \rho(z) \De z \gg 1$ atoms within 
a small interval $\De z$ around position $z$ \cite{EITrev2005}. 
These operators obey the relations 
$\sig_{\mu\nu}(z) \sig_{\nu' \mu'}(z') = \sig_{\mu \mu'}(z) \de_{\nu \nu'} 
\de(z-z')/\rho(z)$. 

Using the above Hamiltonian, while assuming that most of the atoms are 
in the ground state $\ket{g}$ 
(so we can set $\sig_{gg} \simeq \mathds{1}$ and $\sig_{ee},\sig_{es} \to 0$), 
we obtain the following Heisenberg equations of motion for the 
field $\hat{\mc{E}}(z)$, 
atomic polarization $\hat{P}(z)  \equiv \sqrt{\rho(z)} \sig_{ge} (z)$, and 
Rydberg excitation spin-wave $\hat{S}(z)  \equiv \sqrt{\rho(z)} \sig_{gs} (z)$ 
operators,
\begin{eqnarray}
& ( \partial_t + c \partial_z ) \hat{\mc{E}}(z) = 
i g \sqrt{\rho(z)} \hat{P}(z), \label{eq:Ep}  \\
& \partial_t \hat{P}(z) = (i \De_p - \ga_e) \hat{P}(z) 
+ i g \sqrt{\rho(z)} \hat{\mc{E}}(z) + i \Om_c^* \hat{S}(z)  , \qquad 
\label{eq:sigge} \\
& \partial_t \hat{S}(z) = (i \De_p - \ga_s ) \hat{S}(z) + 
 i \Om_c^* \hat{P}(z) ,
\label{eq:siggs} 
\end{eqnarray}
where $\ga_e \gtrsim \Ga_e/2$ and $\ga_s = \Ga_s/2+ \ga_{sg}$ 
are the atomic relaxation rates, and we neglect the associated 
Langevin noise as it does not affect our results 
\cite{Fleischhauer2000,EITrev2005,Gorshkov2007}. 

In the stationary regime, we can solve the above equations in the 
steady state to obtain the field propagation equation 
$\partial_z \hat{\mc{E}}=i\frac{\om_p}{2c} \chi \, \hat{\mc{E}}$ 
with the medium susceptibility
\begin{equation}
\chi(z,\De_p) = \frac{2}{\om_p} \frac{i g^2 \rho(z)}
{\ga_e - i \De_p + \frac{|\Om_c|^2}
{\ga_s - i \De_p} } . \label{eq:chi}
\end{equation}
In the presence of the control field $\Om_c \neq 0$ and small 
$\ga_s \ll w$ (as is the case for the long-lived Rydberg states, 
$\Ga_s \approx n^{-3} \Ga_e$), we have vanishing absorption 
$\frac{\om_p}{2c} \mathrm{Im} \chi(z,0) \to 0$ 
of the EIT medium for the resonant probe field, $\De_p=0$ 
\cite{Fleischhauer2000,EITrev2005,Gorshkov2007}.  
Importantly the EIT bandwidth $w= \frac{|\Om_c|^2}{\ga_e}$ is the same
as that of the single photon resulting from the conversion of the
Rydberg excitation spin-wave by the control field. This is because
the two processes are closely related, as will become apparent shortly.

Consider the adiabatic conversion of excitation between the spin-wave 
$\hat{S}(z,t)$ and field $\hat{\mc{E}}(z,t)$ using the time-dependent 
control field $\Om_c(t)$. For $\De_p=0$ and neglecting the Rydberg
state relaxation $\ga_s$, combining Eqs. (\ref{eq:Ep}) and (\ref{eq:siggs}) 
we obtain $( \partial_t + c \partial_z ) \hat{\mc{E}}(z,t) 
= \frac{g \sqrt{\rho(z)}}{\Om_c(t)} \partial_t \hat{S}(z,t)$.
Following \cite{Fleischhauer2000}, we 
%perform a canonical transformation and 
define the dark-state polariton operator 
$\hat{\Psi} = \cos \Theta \, \hat{\mc{E}} - \sin \Theta \, \hat{S}$ 
which is a coherent superposition of the field and spin-wave operators, 
with the mixing angle defined via $\tan \Theta = \frac{g \sqrt{\rho}}{\Om_c}$. 
%The inverse transformation is given by 
%$\hat{\mc{E}} = \cos \Theta\, \hat{\Psi}$
%and $\hat{S} = - \sin \Theta\, \hat{\Psi}$. 
%This can also be verified using Eq.~(\ref{eq:sigge}) that leads to
%$\hat{S} = - \tan \Theta \, \hat{\mc{E}}$ for the dark state 
%of the three-level system $\ket{D} = 
%\frac{\Om_c \ket{g} - g \mc{E} \ket{s}}{\sqrt{|\Om_c|^2 + g^2 |\mc{E}|^2}}$ 
%that does not contain the excited state $\ket{e}$ 
%\cite{Bergmann1998,EITrev2005,PLDP2007}. 
The propagation equation for the dark-state polariton is
\begin{equation}
\big( \partial_t + v(t)  \partial_z \big) \hat{\Psi} (z,t) = 0 ,
\end{equation}
where $v(t) = c \cos^2 \Theta(t)$ is the group velocity. 
It has a simple solution 
$\hat{\Psi}(z,t) = \hat{\Psi}\big( z - \int_0^t v(t') dt', t=0 \big)$,
which describes a state and shape preserving propagation of the combined
field and atomic spin-wave excitation with the time dependent group velocity 
$v(t)$. For $\Om_c = 0$ ($\cos \Theta =0$) we have the stationary ($v=0$) 
spin-wave $\hat{\Psi} = - \hat{S}$. Turning on the control field $\Om_c$ 
converts the spin-wave into a photon propagating without absorption with 
the velocity $v \to c$ as $\Om_c \gg g \sqrt{\rho}$ ($\cos \Theta =1$).

\subsection{Singe-step creation of the photon}
\label{ap:stirapemit}

We now consider a constant control field $\Omega_c \neq 0$,
and time-dependent excitation field $\Omega(t)$ acting simultaneously 
on the atoms. Combining Eqs. (\ref{eqs:c0cj}) and (\ref{eqs:cjbj}), we have
\begin{subequations}
\label{eqs:c0cjbj}
\begin{eqnarray}
\partial_t c_0 &=& i \sum_{j=1}^N \tilde{D}_j^* c_j , \\
\partial_t c_j &=& i (\tilde{\delta}_j -\Gamma_s /2) c_j
+ i \tilde{D}_j \tilde{c}_0
+ i \Omega_c  e^{i (\bs{k}_c -\bs{k}_0) \cdot \bs{r}_j} b_j , \qquad \\
\partial_t b_j &=& - \hlf \Gamma_e b_j +
i \Omega_c^* e^{i (\bs{k}_0 -\bs{k}_c) \cdot \bs{r}_j} c_j  .
\end{eqnarray}
\end{subequations}
Assuming $|\Omega_c| < \Gamma_e/2$ and small decay and 
two-photon detuning of the Rydberg state
$\Gamma_s,\tilde{\delta}_j \ll w = \frac{|\Omega_c|^2}{\Gamma_e/2}$,
we obtain  
\begin{eqnarray*}
b_j(t) &=& -c_0(t) e^{i (\bs{k}_0 -\bs{k}_c) \cdot \bs{r}_j}
\frac{\tilde{D}_j}{\Omega_c} , \\
c_0(t) &=& c_0(0) e^{- \int_0^t dt' \frac{\bar{D}^2(t')}{w}} .
\end{eqnarray*}
With $c_0(0) =1$, using Eq.~(\ref{eq:akint}) we have
\begin{eqnarray}
a_{\bs{k}} (t) &=& i \tilde{g}_k(t) \sum_j \frac{D_j}{\Delta}
e^{i (\bs{k}_0 - \bs{k}_c -\bs{k}) \cdot \bs{r}_j} , \\
\tilde{g}_k(t) & \equiv & g_k \int_0^t dt'\frac{\Omega(t)}{\Omega_c}
e^{i (\omega_k - \omega_{eg}) t'} e^{- \int_0^t dt'' \frac{\bar{D}^2(t'')}{w}} ,
\nonumber
\end{eqnarray}
which is equivalent to Eq.~(\ref{ap:eq:aksol}) since there
$c_j(0) \propto -D_j/\Delta$.

\subsection{Atomic parameters}
\label{ap:atomparameters}

There are several possible choices of the atoms and their Rydberg states 
to implement our scheme sketched in Fig.~1 of the main text.
One possibility is to use for the source a single trapped Cs atom 
with Rydberg states $\ket{u} \equiv \ket{70P_{3/2},m_j=1/2}$ 
and $\ket{d} \equiv \ket{70S_{1/2},m_j=1/2}$, while the medium atoms
are Rb with Rydberg states $\ket{i} \equiv \ket{58P_{3/2},m_j=3/2}$ 
and $\ket{s} \equiv \ket{57 D_{5/2},m_j=3/2}$, with the quantization
direction along $y$ ($\Delta m_j=0$). Using the quantum defects for 
the corresponding states of Cs and Rb \cite{RydAtoms}, we have
$\Delta_\mr{sa} \equiv \omega_{ud} - \omega_{si} = - 2 \pi \times 94\:$MHz.
Calculation of the transition dipole moments $\wp_{du}$ and $\wp_{si}$
involving the radial \cite{Kaulakys1995} and angular parts, leads to 
the coefficient $C_3 = 11.7\:\mr{GHz}\,\mu\mathrm{m}^3$. We use
these parameters in the numerical simulations yielding the results 
presented in Figs. 2 and 3 of the main text.

Alternatively, both the source and the medium atoms can be of the 
same species, e.g. Rb. One possible choice of Rydberg levels of Rb is: 
$\ket{u} \equiv \ket{70P_{3/2},m_j=3/2}$ 
and $\ket{d} \equiv \ket{68D_{5/2},m_j=3/2}$ for the source, and
$\ket{i} \equiv \ket{64P_{3/2},m_j=1/2}$ 
and $\ket{s} \equiv \ket{65S_{1/2},m_j=1/2}$ for the medium atoms.
This leads to   
$\Delta_\mr{sa} \equiv \omega_{ud} - \omega_{si} = - 2 \pi \times 92\:$MHz
and $C_3 = 16.1\:\mr{GHz}\,\mu\mathrm{m}^3$.

Another choice of Rydberg levels of Rb could be:
$\ket{u} \equiv \ket{71P_{1/2},m_j=1/2}$ 
and $\ket{d} \equiv \ket{69D_{3/2},m_j=1/2}$ for the source, and
$\ket{i} \equiv \ket{65P_{3/2},m_j=1/2}$ 
and $\ket{s} \equiv \ket{66S_{1/2},m_j=1/2}$ for the medium atoms.
This leads to   
$\Delta_\mr{sa} \equiv \omega_{ud} - \omega_{si} = - 2 \pi \times 86\:$MHz
and even stronger dipole-dipole interaction coefficient 
$C_3 = 20.3\:\mr{GHz}\,\mu\mathrm{m}^3$.

\subsection{Multiple atomic excitations and transitions to other states}
\label{ap:opentrans}

In the main text, we considered a rather idealized setup involving
four levels of the medium atoms and assuming that the transition to
state $\ket{s}$ is accompanied by the transition of the source atom
from state $\ket{u}$ to state $\ket{d}$. There may be, however, 
transitions outside this manifold of states. 

Thus, some of the medium atoms may not reach state
$\ket{s}$ because of residual excitation of non-resonant intermediate
state(s) $\ket{i'}$ with Rabi frequency $\Omega'$ and detuning
$\Delta' \gg \Omega'$.  The small probability of exciting an atom 
to a state $\ket{i'}$ is $P_{i'} \simeq \frac{1}{2}|\Omega'/\Delta'|^2$.   
Such atoms do not resonantly couple to the control field $\Omega_c$ 
and will not participate in the collective photon emission on the
transition  $\ket{e} \to \ket{g}$. Yet, the Rydberg states 
$\ket{i'}$ will eventually decay back to the ground state via cascade 
of intermediate states, with a small fraction $P_{eg}$ of the decay paths 
proceeding through the $\ket{e}\rightarrow\ket{g}$ transition and emitting 
a photon that has spectral overlap with the directed single photon. 
The probability of such a spontaneously emitted photon overlapping 
spatially with the collective mode is 
$P_{\Delta\Omega} \simeq \frac{\Delta\Omega}{4\pi}=\frac{1-\cos\Delta\theta}{2}$.
We can therefore estimate the probability of multiple photon emission as 
$P_{>1}\simeq N P_{i'} P_{eg}P_{\Delta\Omega}$. 
Using parameters from Figs. 2 and 3, and assuming $\Omega' \simeq \Omega$
and $\Delta' \gtrsim \Delta$, we have $P_{i'} \lesssim 0.008$, 
$P_{\Delta\Omega}\simeq0.012$, and taking $P_{eg}=0.1$ with $N=1000$ we 
find $P_{>1}\simeq 0.01$, which verifies the single photon character 
of the emission. 

Similar arguments apply to non-resonant excitation of multiple 
atoms to the intermediate state $\ket{i}$ with probability 
$P_{i} \simeq |\Omega/\Delta|^2 \ll 1$: only one, out of all 
$N P_{i}$ atoms, can be promoted to state $\ket{s}$, because the 
transition of the medium atoms from state $\ket{i}$ to state $\ket{s}$ is, 
by necessity, accompanied by the transition of the source atom from 
state $\ket{u}$ to state $\ket{d}$, 
as the latter has to supply the microwave photon energy $\omega_{ud}$.
The success of this process can be verified by detecting the source atom
in state $\ket{d}$. If, however, the medium atoms are transferred to some
other state $\ket{s'}$, and correspondingly the source atom to another
state  $\ket{d'}$, it can also be detected by the same measurement.

We finally stress that multiple excitations of the medium atoms to state 
$\ket{s}$ (or $\ket{s'}$) are impossible, if initially there is only 
one source atom in state $\ket{u}$, while even a single Rydberg excitation 
to state $\ket{s}$ will not take place (and the optical photon will not 
be emitted) if there is initially no source atom in state $\ket{u}$.

\end{document}